# Trading green bonds using distributed ledger technology

*Pre-print*

Henrik Axelsen, University of Copenhagen, Denmark, heax@di.ku.dk

Ulrik Rasmussen, Deon Digital, Denmark, ulrik.rasmussen@deondigital.com

Johannes Rude Jensen, University of Copenhagen and eToro Labs, Denmark, j.jensen@di.ku.dk

Omri Ross, University of Copenhagen and eToro Labs, Denmark, omri@di.ku.dk

Fritz Henglein, University of Copenhagen and Deon Digital, Denmark, henglein@diku.dk

## Abstract

*The promising markets for voluntary carbon credits are faced with crippling challenges to the certification of carbon sequestration and the lack of scalable market infrastructure in which companies and institutions can invest in carbon offsetting. This amounts to a funding problem for green transition projects, such as in the agricultural sector, since farmers need access to the liquidity needed to fund the transition to sustainable practices. We explore the feasibility of mitigating infrastructural challenges based on a DLT Trading and Settlement System for green bonds. The artefact employs a multi-sharded architecture in which the nodes retain carefully orchestrated responsibilities in the functioning of the network. We evaluate the artefact in a supranational context with an EU-based regulator as part of a regulatory sandbox program targeting the new EU DLT Pilot regime. By conducting design-driven research with stakeholders from industrial and governmental bodies, we contribute to the IS literature on the practical implications of DLT.*

*Keywords: DLT Pilot regime, Trading, Settlement, Liquidity, Green Bonds, Net-Zero, Funding*

# 1   Introduction

To achieve the UN climate conference (COP21) Paris Agreement of limiting global warming to less than 2.0 degrees Celsius and providing a significant effort to limit it to 1.5 degrees Celsius this century, the international community must reach net zero carbon emissions by 2050 (United Nations, 2015). According to global observers, the planned energy scenario to reduce these required carbon emissions may require an investment of up to USD 95 trillion (IRENA, 2020) from 2016 to 2050, with more transformative scenarios requiring even more.

Voluntary carbon markets (VCM) play a key role in this transition. Voluntary markets differ from the emissions-based carbon credit markets (so-called compliance carbon markets) by enabling the trade of carbon sequestration, avoidance of nature loss, and other efforts to reduce carbon emissions, including technological improvements. Although VCM has shown impressive growth over the past decade, the concept is plagued by two key problems: Correctly certifying the integrity of the carbon credit presents an extraordinary challenge, and the lack of a globally scalable and compliant trading infrastructure greatly limits the issuance and trade of these instruments.

This paper focuses on the second problem: Designing a scalable financial market infrastructure for the voluntary carbon markets. Both problems are intimately connected, as carbon credits' perceived and real value depends on the verifiable integrity of the underlying carbon capture certificates. We set out to explore the following research question: '*To what extent can distributed ledger technology (DLT) facilitate the issuance, trading, and settlement of regulated financial instruments (green bonds) to finance carbon capture based on verified carbon credits in voluntary carbon markets*?'

Our research question naturally implies a need to understand the implications of financial securities regulation. For this reason, we worked with a team consisting of financial, agricultural, and technological experts. Some of us entered a regulatory sandbox program under the EU DLT Pilot regime, led by a Financial Supervisory Authority from an EU member state. The program aimed to assess the regulatory requirements of a novel DLT-based Trading and Settlement System (DLT TSS) based on domain-specific language technology for specifying financial and commercial contracts (Andersen *et al.*, 2006).

By conducting multidisciplinary and design-driven research, we contribute to the Information Systems (IS) discourse on the practical implications and limitations of DLT. This involves a novel DLT-based artefact addressing the need to scale voluntary carbon markets to a global group of users. We examine how DLT infrastructure may be used to scale VCM to introduce these instruments into existing trading and settlement systems. We use the novel concept of 'carbon cash flows' as collateral, originated via projects, to demonstrate the benefits of using DLT as capital market infrastructure. Defending market integrity and stability is critical to designing and evaluating a capital market infrastructure. An equally important, but distinctly different challenge, is the complexity of environmental market integrity, which we will touch only briefly in this paper as it relates to financial market integrity. The market infrastructure artefact presented here facilitates the monitoring and reporting of such strict verification and integrity standards as how marketplaces in traditional capital markets operate.

We use the Design Science Research (DSR) methodology, informing an iterative approach to the artefact design in which feedback from stakeholders representing multidisciplinary perspectives is integrated into the design specification. To this extent we aim to contribute practical insights to the growing body of IS literature demonstrating the application of the DSR methodology in the design of technical artefacts addressing the challenges of today.

## 2      Background

IS scholars have long promoted technologies broadly referred to under the 'DLT' umbrella for the benefits that these may imbue on regulated capital markets (Collomb and Sok, 2016) and payments in general (Lindman, Rossi and Tuunainen, 2017). The potential of DLT and Blockchain in pre- and post-trade processes is well examined in the literature, suggesting a potential for reducing costs while mitigating counterparty credit risk (Jensen and Ross, 2021) and reducing the cost of capital. Several other efficiency gains have been identified in the literature, ranging from transparency in the verification of securities holdings, mutualization data, and optimized Know-Your-Customer (KYC) processes (Parra-Moyano and Ross, 2017) in pre-trading to real-time transaction matching, execution, and reporting. The IS literature frequently uses design-oriented or case-based methodologies to explore and demonstrate how new DLT relieves or creates friction across industries. Scholars have shown how blockchain might give rise to new types of economic systems (Beck and Müller-bloch, 2018), or how the implementation of blockchain technology introduces fascinating organizational issues (Gozman, Liebenau and Aste, 2020).

## 2.1 The Markets for Voluntary Carbon Credits (VCC)

The voluntary carbon market differs from the general carbon compliance markets for designated carbon offsets, associated with the international efforts led by the United Nations Climate Change Convention. VCMs let developers of projects that prevent, reduce, or eliminate carbon emissions apply to private standardization organizations, which then certify the emissions avoided, reduced, or eliminated by the project. Developers create voluntary carbon credits (VCCs) through a designated certification process in which one VCC represents one ton of $CO_2$ emission captured or avoided. The VCCs are stored in a registry maintained by the organization that certifies the project. To claim the reductions, the developer can either retire the credits to offset $CO_2$ emissions or transfer them to another organization with an account in the registry.

In simple terms, the business case for VCCs is to unlock funding for those willing to commit to the preservation of the cultivation of forests or other events leading to the increased sequestration of carbon from the atmosphere. Let us consider an example: A small farmer is looking to transition from current farming methods to regenerative farming methods. To do so, the farmer will need to acquire new machinery and other types of seed, which will introduce several new expenses. The new regenerative farming methods will typically result in reduced crop yield for a few years before producing results like non-regenerative methods. As a consequence of the transition, the farmer will face increased costs and new risks to her existing revenue streams. By sourcing new revenue streams through the sale of carbon credits, the farmer can make up for the shortfall over time. Indeed, bridging this liquidity gap is in the global community's shared interest, as the lack of financial incentives is a major obstacle to accelerating the transition to Net Zero. Yet, because of the issues outlined above, small and medium enterprises (SME) are disincentivized from pursuing a green transition. Collateralization of future green cash flows through so-called Asset-Backed Securities (ABS) is gaining prominence but has yet to reach SMEs (Global Capital, 2022).

In recent years, global initiatives such as the Taskforce for Scaling Voluntary Carbon Markets (TSVCM, 2021) have been mandated to accelerate growth in these markets. In addition to proposing integrity principles for voluntary carbon markets, the TSVCM and now its successor, the ICVCM, suggests that new infrastructure is needed to provide the backbone for trading, clearing, and settlement of VCC, coupled with new funding solutions that can produce transparent market and reference data. The suggestions emphasize meeting the increasing supply and demand for VCC by building (1) exchanges that will manage Core Carbon Principle aligned credits to enable increased liquidity and ease of purchase, (2) post-trade infrastructure, including the design and supervision of a meta-registry to bolster market integrity and market functioning, and (3) advanced and transparent data infrastructure with shared protocols that are widely accessible.

## 2.2  Is Blockchain Technology the Solution?

Recent years have seen several attempts at using blockchain technology for VCC trading (Dodge, 2018). Proponents of the concept argue that the technology has the potential to improve liquidity while reducing transaction costs (Kotsialou, Kuralbayeva and Laing, 2022). Several of these attempts have come from the "wild west" of Decentralized Finance (DeFi) (Sipthorpe *et al.*, 2022) integrating novel concepts such as NFTs and stablecoins under the moniker of regenerative finance (ReFi). Unfortunately, leading projects have been hit by a slew of scandals related to questionable approaches to the qualities referred to as permanence, leakage, and additionality in the VCM literature.

Additionality refers to the principle that only carbon capture or emission avoidance that would otherwise not have happened by itself can be awarded carbon credits. Leakage refers to the problem of carbon emissions being moved from a carbon capture project area to another area, such as cutting down another forest instead of the one entering the project. Permanence refers to the principle that carbon capture or emission avoidance must effectively last forever to be valid: Capturing carbon (while receiving credits for it) and subsequently releasing it again (without repaying the carbon credits) has no net carbon capture effect.

As a result of questionable practices and doubts about permanence, the leading VCC verification agency Verra suspended verification for tokenized credits traded in DeFi applications in the spring of 2022 (Ledger Insights - blockchain for enterprise, 2022). This decision was made due to potential fraud in the retirement of tokens and the risk of double spending, which questions the overall integrity of the markets. Verra currently verifies almost two-thirds of all VCCs and has recently launched a consultation process to investigate how to create the required integrity for VCCs issued on public chains. Despite these temporary setbacks, it has become increasingly clear that the transparency and tracking capabilities associated with DLT and blockchain provide an interesting opportunity for bootstrapping VCM, including using a shared digital data protocol across the voluntary carbon standards to improve speed, accuracy, and data integrity.

## 2.3 Regulation

The EU Commission's digital finance agenda in 2020 delivered several groundbreaking regulations in 2022, namely the DLT Pilot Regime regulation no. 2022/858 ('DLTR') coming into effect in 2023, and the Markets in Crypto Assets regulation ('MiCA') coming into force likely in 2024. MiCA and DLTR use the same definition of DLT but approach the topic from different jurisdictions. MiCA will regulate crypto assets that are not securities. Tokenized securities, i.e., digital representations of existing securities, are regulated by the existing securities regulations, namely the Market in Financial Instruments Directive II and associated regulation (MiFID/MiFIR) as well as the Central Securities Depository Regulation (CSDR), and DLTR. The DLTR provides a potential means to use DLT for trading and settlement systems under those regulations with appropriate exemptions from the regulatory playing field required due to the DLT-based execution. The assessment of suitability takes place in a so-called sandbox, a new regulatory invention, providing a supervisory environment where representatives from the authorities participate in a process of knowledge exchange on novel technology, in exchange for assessment and guidance on eventual licensing and regulatory integration.

## 3 Method

We utilize the Design Science Research (DSR) methodology to form an iterative process (Gregor and Hevner, 2013) in which new versions of the artefact are designed and then presented to stakeholders for feedback. Each cycle seeks to integrate the increasingly expansive list of artefact requirements emerging throughout the design-search process. The overall project spanned a duration of 14 months, in which the team, including the authors, students providing prototype system implementations, and project partners conceptualized and designed the artefact by implementing variations of the following 6-step process, drawn from the DSR literature: 1) Problem identification, 2) Solution objective, 3) Design, 4) Demonstration, 5) Evaluation and 6) Communication (Peffers *et al.*, 2007). In its final phase, the project involved a group of eight external experts alongside the authors, who participated actively in the design-search process throughout the project's duration (Table 1).

| #  | Role in host-organization | Role in the design search process |
|----|---------------------------|-----------------------------------|
| S1 | Special Advisor, Banking Technology Company | Domain expertise, requirements design and evaluation, guidance, support |
| S2 | Partner, regulatory consulting firm | Non-functional requirements design and evaluation |
| S3 | Developer, technology startup | Design, test, implementation, and evaluation of functional requirements |
| S4 | Project lead, Agtech startup | Voluntary carbon markets domain expertise |
| S5 | Capital markets expert, Regulator | Requirements design Multilateral Trade Facility (MTF), artefact evaluation |
| S6 | Fintech expert, Regulator | Domain expertise, guidance, and support |
| S7 | Fintech expert, Regulator | Domain expertise, guidance, and support |
| S8 | Capital markets expert, Regulator | Requirements design CSDR and Regulated Markets (RM), artefact evaluation |

*Table 1. Stakeholder categories and role in the search process*

The project was developed over three distinct phases:

In the first phase, we delivered a conceptual demonstration of the artefact to facilitate discussion while identifying and engaging stakeholders that would help demystify the technological challenges within the regulatory context.

The second phase introduced a Proof of Concept (POC) for the artefact, demonstrating a traditional order book and delivery-versus-payment (DvP) settlement system. The POC was presented to a broader audience in a financial incubator alongside colleagues inside and outside of IS to gauge interest in the concept and collect early feedback from peers. During the second phase of our process, it became clear that new regulation on DLT was to be implemented at the EU level, which would come to present a much clearer regulatory environment for the artefact. These developments led us to consider whether the artefact could be a potential candidate to support the scaling of VCM at the EU level. For this reason, we sought access to a regulatory sandbox with an EU-based National Competent Authority (NCA), positioning the artefact as a potential accelerator for funding the liquidity gaps and frictions related to the securitization of sustainable funding for SMEs. The search was successful, leading to the development of a pre-production level version of the artefact required for a formal assessment of compliance in collaboration with representatives from the NCA. Due to the restricted scope of the (pilot) DLTR regulation, we re-designed the approach to fit traditional financial instruments, approaching the VCM project funding challenge with a bond structure, iterating away from the initially targeted classical trading of VCC certificates. This development, in turn, changed the focus of the design work from mainly being around the complexities of environmental integrity towards a primary focus on the complexities of financial integrity in a regulated capital market infrastructure.

The third phase introduced an element of intensive regulatory scrutiny, emphasizing and challenging the rationale for existing securities regulation. The intent was to explore reasonable exemptions from existing regulations considering the forthcoming DLTR. By working with representatives directly involved in the negotiation of forthcoming regulation, the team was able to align the artefact for compliance with regulation coming into force in 2023.

The artefact evaluation was conducted ex-ante through expert interviews within the context of the confidential regulatory sandbox (Venable, Pries-Heje and Baskerville, 2016). The evaluation sessions generally took the form of technical demonstrations. As the design-search process progressed, the format was advanced to feature workshop presentations in which the artefact was put to work by demonstrating test scripts. A general model for the design-search process is outlined in figure 1 below.

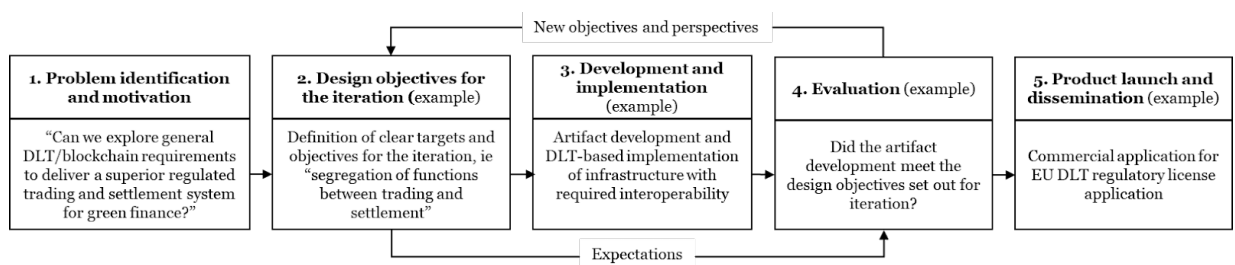

*Figure 1. The DSR method applied to the project phases*

### 3.1   Artefact Requirements

Through several iterations, the group of stakeholders delineated a set of requirements for the final iteration of the artefact. In Table 2 we feature a summarized version of the requirements for the latest cycle.

| Category | Details |
|---|---|
| Core Technical Requirements | (1) Manage states of contracts across the securities' lifecycle.<br>(2) Identify and verify that users are authorized for their roles.<br>(3) Maintain ownership of securities<br>(4) Guarantee atomic, consistent, isolated, and durable (ACID) execution of compound transactions, specifically delivery versus payment.<br>(5) Enforce correct attribution and non-repudiability of actions (using digital signatures and cryptographic commitments). |
| Contextual Requirements and Objectives | (1) Interoperability with external systems.<br>(2) Settlement finality: The determination of a definite time after which the transfer of legal title (ownership) is irrevocable.<br>(3) Support for new financial instruments with high-frequency data dependencies (e.g., carbon emission monitoring data).<br>(4) DLTR compliance with well-reasoned exemptions from existing regulations written for traditional centralized systems.<br>(5) Interoperability with legacy private and central banking as well as private, permissioned, and permissionless DLT/blockchain and other clearing and settlement systems.<br>(6) Support for full access by the financial supervisor/regulator to maximize automated supervision.<br>(7) Full transparency and traceability of underlying verification data throughout carbon credit and advanced instruments' lifecycle.<br>(8) Efficient high-volume trading processing, instantaneous settlement (execution) of trades, real-time monitoring, and advanced market abuse detection<br>(9) Ability to catalyze structured finance by domain-specific language for specifying new instruments and immediately issuing them. |

*Table 2. Functional Artefact Requirements*

# 4 Artefact Description

The designed artefact is based on the Smart Financial Instrument (SFI) system (Deon Digital, 2023), a smart contract platform developed by Deon Digital with the express purpose of servicing the lifecycle of both regulated and unregulated digital assets and securities. In the present case, the design focuses on the trade and execution of voluntary carbon credits with emphasis on being a regulated tradeable security. Figure 3 displays the high-level architecture and subsystems of the artefact. Their responsibilities (functionality) are listed in Table 3.

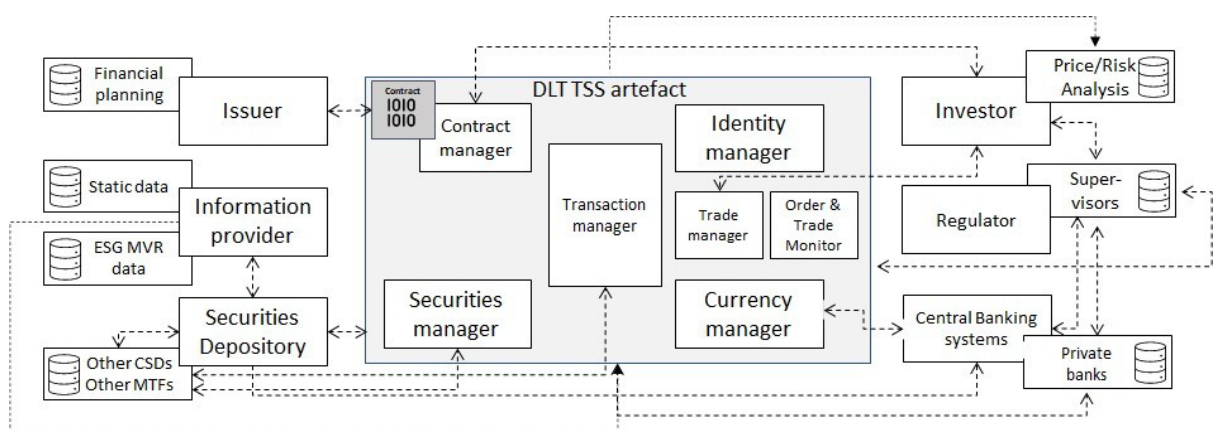

*Figure 3: Overview of the artefact architecture and subsystems*

The artefact aims at improving the technological shortcomings and reconciliation issues in existing capital market infrastructure by consolidating and formalizing all data, interpretation, and logic necessary to express the nature and lifecycle of the financial instrument.

| Identity manager | Maintains up-to-date information about users authorized to access the artefact. This includes regularly updated information sufficient to satisfy KYC/AML/CFT requirements including public keys registered by users for verification of their digitally signed messages. |
|---|---|
| Contract manager | Receives legally required instrument documents (prospectus and term sheet) and formal contract specification of financial instrument, gets financial instrument approved and rated by designated outside services, launches the instrument so it is live, and checks and processes life-cycle events such as payment instructions and notifications according to formal contract specification of the instrument. |

| Security manager | Manages ownership of issued instruments reservations (to support transactional exchanges), collateralization (a form of reservation), and finalized transfer of title (full ownership transfer). |
|---|---|
| Currency manager | Manages ownership of fiat currencies, in particular reservations (to support atomic exchanges, specifically delivery-versus-payment), collateralization, and finalized currency transfers. |
| Trade manager | Manages matching of buy and sell orders of instruments and their immediate settlement; in particular, reserves currency (buy offer)/instruments (sell offer) when receiving an offer, performs matching of buy and sell offers resulting in a spot exchange contract (trade), and settles the trade in real time via atomic transfers in the security and currency managers. |
| Transaction manager | Stateless service that guarantees logical atomicity, all-or-nothing execution of a set of state changes in multiple state managers, e.g. to guarantee atomic delivery versus payment. |
| Order and trade monitor | Configurable, stateless service that subscribes to the trade manager to receive both authenticated orders and settled trades and performs both real-time and ex-post analysis of orders and trades. It automatically identifies single or connected groups of orders and trades as suspicious based on configurable specifications and market abuse detection techniques. |

*Table 3. Overview of architecture components.*

### 4.1 State managers

The first five of the subsystems in Table 3 are state managers. In general, a state manager is characterized by a mathematical function f, that deterministically computes its current state *s* from its current log *l* of previously validated events; that is, $s = f(l)$. If $l' = l*e$ (*l* extended with a new event *e*), then the state is updated to $s' = f(l')$ using an efficient incremental-update version of *f*. In particular, the unique correct state of a state manager can be reconstructed after a crash failure and checked for correctness at any time by an external service from the mathematical definition of *f* and the tamper-evident ledger of previously validated events. A state manager provides an API for submitting events for validation, querying the current state, subscribing to new validated events and supporting 2-phase commit for synchronized commitment of multiple events and their storage on multiple ledgers.

### 4.1.1   Resource managers

The currency and security managers are examples of resource managers. A resource manager maintains ownership and processes transfers of any number of resource types (currencies, assets, tokens, etc). They guarantee that the total amount of resources in the system is constant and that transfers can be performed in any order that does not violate their owners' credit limits. In particular, this means that the question of enforcing zero credit limits, corresponding to balances of ordinary users not being allowed to become negative, is the only ``real'' consensus problem requiring more than point-to-point communication between authenticated agents (Henglein, 2018).

The security manager maintains the balance of ownership of the securities it manages. The function $f$ in this case, is the summing of the validated transfers (viewed as a suitable mathematical structure) in the ledger. The balance is used in validating submitted transfer instructions: a transfer that would result in a negative balance is rejected by the security manager.

The currency manager maintains fiat currency accounts if its operator is licensed to do so. Alternatively, it is implemented as a proxy service to a banking or e-money institute API where the accounts are held. Likewise, it can also be implemented as a proxy service for blockchain systems if payments are to be made in stablecoins.

### 4.1.2   Contract managers

A contract manager maintains the authoritative state of a set of issued financial instruments that are still live. It is a state machine that maintains the current state $c$ of a financial instrument identified by International Securities Identification Number (ISIN) number $I$ according to the instrument's formal contract specification in the domain specific language CSL (CSL Platform Documentation). A contract manager receives a digitally signed event $e$, for example a coupon payment instruction, for $c$, one of the instruments it is in charge of, from a client and checks whether $e$ matches $c$ in the sense of being an admissible action according to the contract specification. After validation by other managers, if any (for example, the currency manager executing/validating the payment instruction), it logs $e$ associated with $I$ in its ledger and informs subscribers (clients) of this event having happened. It also updates the state of $I$ from $s$, before the coupon payment, to its new residual state $s'$, the state of $I$ after the coupon payment. Clients can query the authoritative state of $I$ and may submit bids and offers for $I$ tied to a particular state to ensure that an offer to buy $I$ in the state before a coupon payment is not matched with an offer to sell $I$ after coupon payment.

Contract managers do not require synchronization amongst themselves since the order of events for different contracts is *a priori* irrelevant; they are only synchronized via a resource manager in case of a resource transfer. For example, a notification by the issuer to execute a prepayment clause of a bond requires no synchronization with any other events and thus no communication with other contract or resource managers.

## 4.2 Transaction managers and network activity monitors

A transaction manager is an essentially stateless service that effects atomic transactions, that is all-or-nothing updates of multiple state managers using a customized 2-phase distributed commit protocol.

They only require local state during a transaction, which does not need to be retained once a transaction is concluded. Consequently, any number of independent network nodes, each running an independent transaction manager, can be employed for scalability.

An order and trade monitor monitors suspicious trading activities as required by regulation. Any such activity is then filtered by tool-supported human analysis for eventual regulatory reporting to the supervisory authority. Additionally, it provides an API for the supervisory authority to submit and execute queries/programs of their own choosing on the order and trade data that are securely and authoritatively logged in the trade manager ledger. Multiple monitors operating on independent network nodes, each monitoring a distinct set of instruments, can be employed for scalability.

This demonstrates the feasibility of a modularized internal market surveillance function that operates by subscribing to the trade manager's received and validated orders as well as (settled) trades. This facilitates 'embedded supervision,' where the NCA is authorized to install their own fully automated trade supervision modules as a regulatory observer rather than engaging in lengthy and mutually costly email interchanges requesting certain data in (imprecise) natural language (Axelsen, Jensen and Ross, 2022).

## 4.3    Distributed Systems Architecture

The artefact employs a two-level distributed systems architecture. At the top level it consists of independent state managers with distinct functionalities, each of which has its own thread of control and maintains its own cryptographically secured append-only digital ledger. The state managers are coordinated by transaction managers employing a distributed 2-phase commit protocol to ensure atomic execution of multiple actions. At the bottom level, each of these subsystems is implemented by a small set of nodes employing an active replication protocol for crash failure resistance.

In a simplified implementation there are no secondaries in the bottom layer: Each state manager is implemented as a single node on a secure network whose ledger is continuously written to local disk storage and, in encrypted form, to an off-site secure storage facility. This constitutes a permissioned DLT-based system: All messages are digitally signed where every user and node operator is identifiable as a legal person by the identity manager. In particular, the digital signature in a message, for example a payment instruction, provides non-repudiable evidence that (somebody having access to the private keys of) a specific, identified legal entity has authored the message.

The artefact is functionally sharded: It has no global blockchain that sequences all recorded events whether doing so is logically actually required. The conceptually collective state of the ledgers *in toto* comprises the set of all validated state messages of all state managers. They are synchronized across state managers to the degree logically necessary.

Note that this is different from both mainframe systems and conventional blockchain systems, whether permissioned or non-permissioned, where arriving messages are sorted into a single linear stream of events prior to processing them, irrespective of whether such ordering is needed in an application. Consensus on a total order of events arriving at the network nodes of a distributed system, however, is an inherently severe performance bottleneck for any distributed system, including both non-permissioned and permissioned blockchain systems. It is ultimately even unsolvable in deterministic asynchronous distributed systems with just one node that can fail (Fischer, Lynch and Paterson, 1985), which expresses itself as the ``trilemma'' problem in blockchain systems.

The relatively easy programming problem of writing reactive single-threaded code in an Ethereum-style blockchain system is thus bought at the great expense of solving an inherently hard problem at each step: distributed consensus by all nodes on a specific order of a block of events. In contrast, the artefact does not build a global blockchain or any other data structure implementing a single linear sequence of events. Its `functional' sharding yields scalability: Ownership of securities and money is managed in independent subsystems whose execution is not fine-grained synchronized. Instead, the artefact's transaction manager synchronizes updates on multiple state managers only when needed.

## 4.4 VCC Instrument Execution

An issuer can issue a bond whose life-cycle actions include not only payments to the investors but also information events provided by designated verification assurance and calculation agents. This is described by the following state changes, visualized as a UML sequence diagram in figure 4.

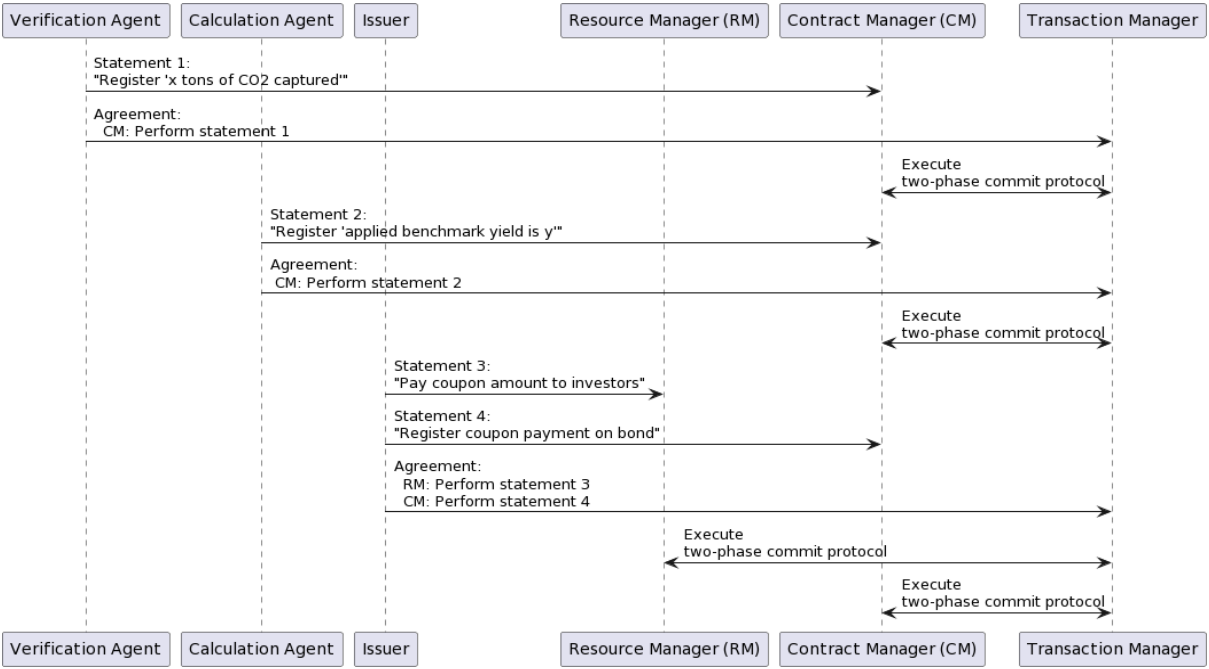

*Figure 4. Sequence diagram for bond with coupon payments dependent on data from independent agents*

There are four statements executed among 6 participants (nodes): (1) 'Verification Agent' V verifies to 'Contract Manager' CM that X ton of $CO_2$ has been captured within the scope of the given green bond. V also confirms to 'Transaction Manager' TM to execute this statement as a two-phase commit. (2) 'Calculation Agent' C confirms to CM to register the yield as Y, while also instructing TM to perform this statement, again, as one two-phase commit. (3) 'Issuer' I then instruct 'Resource Manager' RM to pay this coupon to investors, and (4) CM to register that coupon payment on the bond. Once confirmed, the TM executes those instructions, again with a two-phase commit.

## 5     Results and Evaluation

The final evaluation was conducted with the full panel of stakeholders representing industry and regulators and uncovered several interesting perspectives on the feasibility of the artefact. First, we examine the core technical requirements posed to the artefact (Table 4).

| # | Requirement | Evaluation summary |
|---|---|---|
| 1 | Manage states of contracts | The artefact demonstrates both conceptually and in regulatory testing how states of contracts are managed during all steps of a financial instrument's life cycle according to its formal contract specification. |
| 2 | Identify and verify users | The artefact demonstrates how identities are established, verified, and authorized for their role. |
| 3 | Maintain ownership | The artefact demonstrates how ownership of securities, monies, and other assets are safely and transparently maintained using resource managers that guarantee that resources can neither be digitally lost nor duplicated. |
| 4 | Guaranteed transactionality | The artefact guarantees atomic execution of compound transactions across internal and external subsystems, in particular delivery versus payment. |
| 5 | Non-repudiability | The artefact maintains a tamper-evident, securely stored ledger of the authoritative sequence of events, each non-repudiably digitally signed by legally identified agents. |

*Table 4. Evaluation of core technical requirements.*

## 5.1 Throughput, Finality, Interoperability, and Settlement

The attainable throughput for the artefact was tested with a hand-coded complex financial instrument with several life-cycle activities to evaluate its throughput potential. Results demonstrated throughput between 200,000-13,000,000 events per second on a single standard cloud-hosted server, depending on how the digital signature checking was implemented and how often checking signatures was needed (Petersen, et al., 2022). The events included both price observations (predominantly) and payments. Compared with the Nasdaq Historical TotalView-ITCH requirements, which contains all events in every instrument traded on the Nasdaq exchange, a standard limit order book structured exchange handles up to 200,000 messages per second. A conservative fully distributed implementation of the artefact can be expected to reach more than 1,000,000 messages per second even with full individual digital signature checking since total-event-order consensus across different financial instruments being life-cycled and traded is neither needed nor implemented.

In a theoretical near-perfect implementation with high-performance computing infrastructure distributed across thousands of servers, the artefact may even scale beyond these numbers. Yet, high processing throughput does not necessarily translate into the ability to validate transactions fast in practice, as there are several real-world drivers of potential latency, especially in identity management if authentication involves external services and settlement of payment instructions involving central or commercial bank money.

A key requirement for securities settlement systems, such as T2S at the European Central Bank, is to comply with the EU settlement finality directive, which stipulates that there be a well-defined point in time after which transfers are irrevocable to secure the rights of creditors after a transferor's default. To date, there are concerns about the extent to which blockchain systems employing probabilistic consensus meet the 'deterministic eligibility criteria' (ECB Advisory Groups on Market Infrastructures for Securities and Collateral and for Payments, 2021) for Delivery vs. Payment (DvP).

While IS scholars generally accept blockchain finality as sufficient for verifying the integrity of a transaction once a sufficient number of blocks have been verified (Nærland and Müller-bloch, 2017) concerned voices at regulatory institutions do not approve of probabilistic settlement finality. The artefact provides deterministic finality of security transfers via its own security manager and irrevocability of payment instructions issued to the banking system. Its transaction manager can even act as a real-time bridge with deterministic finality between multiple external systems including blockchain systems if these provide support for reservations (precommit) and subsequently releasing (commit) or returning (abort) reserved resources deterministically. The former is easily programmable as smart contracts; the latter, however, is problematic for Ethereum-style permissionless systems employing probabilistic consensus, which is either slow (takes more than a couple of seconds) or risks retraction (previously confirmed transfers are implicitly revoked when a longer chain without it appears) or both. While the artefact is capable of interoperating with permissionless blockchain systems via its transaction manager, its finality and regulatory acceptance depends on the finality of participating systems.

In the words of DLTR, the 'Union financial services legislation was not designed with distributed ledger technology and crypto-assets in mind and contains provisions that potentially preclude or limit the use of distributed ledger technology in the issuance, trading, and settlement of crypto-assets that qualify as financial instruments.' DLTR defines a 'distributed ledger' as an information repository maintaining records of transactions synchronized between network nodes using a consensus mechanism. Curiously, DLTR does not explicitly require tamper-evident or, stronger yet, tamper-proof recording, which is usually taken to be a core characteristic of blockchain/DLT systems ( Henglein, 2018, Kolb *et al.*, 2020).

The artefact adequately meets the definition set out in the DLTR package: Instruments are formally specified, and the artefact maintains their definitive, unambiguous current state throughout their lifecycle. Since settlement is instantaneous, there is no settlement risk, except for any latency added in the payment settlement leg if that is done in fiat currencies such as EUR or USD through the banking system. In particular, no central counterparty is required to protect trading partners from their counterparties' inability or unwillingness to deliver on their part of the bargain. All information can be provided on equal terms to all users. Bids, offers, and trades are digitally signed, processed, matched, settled, and securely stored in seconds ('T+0') rather than days ('T+2'), the current standard in traditional capital market infrastructure. The current and all previous states and all events relevant to an instrument can be inspected and independently verified based on the contract manager's immutable ledger and the instrument's formal specification. The artefact provides crash-fault tolerance, and any state manager that does not implement its semantics correctly is discoverable and is treated as failed by all other (non-Byzantine) managers.

In the light of these features, the artefact may qualify for exemptions to regulatory requirements disallowing direct retail participation, as there is no settlement risk and hence no direct insolvency risk, and the artefact does not require the obligatory traditional custody and servicing of assets by banks and brokers.

## 6 Discussion

Voluntary carbon markets (VCMs) are currently maintained in a way that separates the registry, project documentation, carbon credit documentation, and trading contract. When a registry issues a Verified Carbon Certificate (VCC) the purchaser must trust that the documentation has been properly examined in accordance with the measurement, verification, and reporting (MVR) protocol. The issuer pays the registry, and an end user of the VCC must trust that this process is accurate and free of any integrity issues as in traditional capital markets, where an issuer pays a rating agency in a similar manner. Furthermore, the prevailing trading contracts of current VCMs refer to a carbon certificate, which is essentially a data record on the registry and must be changed to the new owner or retired manually through a web interface (such as Verra), leading to compartmentalized and isolated information.

The artefact presented here is built on a fully digital representation of the financial instrument with all transactions and evidence recorded immutably, and to the extent this recording includes all previous records, the artefact will enable full transparency of the underlying certificates of the VCM. By creating a one-stop shop with the functionality outlined for the presented artefact, a purchaser has access to all up-to-date information in one place. The platform functionality may also allow an issuer to issue compliant financial instruments that pool credits from similar activities and thus inherently diversify the source of credits and their risks. Traditional VCM markets cannot include such information since the existing capital markets rely on old messaging technology supporting only payments (SWIFT). So, although the artefact design presents significant improvement to the market infrastructure, enabling full documentation verification, protocol compliance, and transparent credit pooling, there is still a level of fragmentation as long as environmental and financial integrity standards are not aligned. As pointed out by TSVCM the long-term solution is integration, which will only happen, when Article 6 of the Paris Agreement is fully completed.

In this paper, we report on the design of an artefact under a new regulatory regime with a group of industrial and regulatory stakeholders. The project was designed to address the research question: '*To which extent can distributed ledger technology (DLT) facilitate efficient issuance, trading and settlement of regulated financial instruments (green bonds) to finance carbon capture based on verified and traceable carbon credits in voluntary carbon markets?*'

The final design of the artefact demonstrates the feasibility of a trading and settlement system for green bonds by satisfying the core technical requirements posed for traditional trading and settlement systems with DLT and formalized contracts for end-to-end digitalization. In addition, the use of DLT introduces several appealing features for the trade and settlement of securities, such as atomic settlement with pre-funded trades, as well as reducing counterparty and liquidity risk in existing T+2 settlement systems, a point frequently raised in the literature (Jensen and Ross, 2021). Compared to conventional blockchain systems, both permissionless and permissioned, that implement a replicated state machine and enforce global consensus amongst all nodes on a particular order of events, the artefact exploits the lack of need for synchronizing *all* events with each other. Synchronizing all events is a built-in bottleneck in blockchain systems, which are required to achieve consensus on a *single* global chain (total order) of all transactions.

Here we have only addressed the issue of VCM infrastructure. To meet the full infrastructure capabilities outlined by TSVCM, the artefact should support the VCC certification process throughout the instrument's lifecycle. The artefact will need to be complemented with advanced analytics add-ons, which can rely on the verified state managers' logs a their single, authenticated source of truth without having to be built into the system itself. Its security registry needs to be implemented as a meta-registry, that is as a proxy service aggregating the collection of individual source registries managed by carbon certificate verification agencies. While the artefact meets the technical requirements identified in the design process, the challenges of integrating DLT-based solutions into the existing financial IT infrastructure remain pertinent. Further, it may also be argued that some regulatory objections to DLT-based solutions in finance are based on technically unwarranted preconceptions and traditions. Addressing these objections should be considered a natural part of the gradual integration of radically innovative technologies (Beck and Müller-Bloch, 2017). As such, it is incumbent upon IS scholars and practitioners to motivate exemptions from traditional securities regulation by showcasing how DLT-based solutions can either reduce frictions in existing markets or enable the flow of funds to otherwise underserved constituents of the financial system. In this paper, we argue that introducing a regulated DLT-based solution in the VCM may incentivize the issuance of VCC-backed securities and promote environmentally sustainable practices in agriculture and beyond, provided that the on-the-ground certification challenges can be overcome.

Summarizing the benefits, challenges, consequences, and mitigation practices required for the adoption of DLT TSS for a Green Bond market, DLT can reduce friction in the lifecycle of financial instruments by executing processes normally requiring multiple service providers within a single component-based distributed architecture.

The transparency of blockchain and DLT-based systems and their novel technology-specific capabilities may be perceived as radical innovations by traditional supervisors, who are used to the standard organization and centralized IT architecture implicitly assumed in current financial regulation. They may question the rationale for allowing distributed systems, mathematically guaranteed transparency, end-to-end digitalization and providing investors with direct control of their assets into financial markets. By gradually designing DLT-based systems to prove value to regulators as guardians of society at large in terms of increased investor protection, market integrity, transparency, and efficiency as well as financial stability, DLT should gain traction considering its superior functionality, transparency and security. But while DLT may be part of the answer, agreements on carbon integrity and financial market integrity principles are required to fully develop a trustworthy sustainable capital market. A secure digital currency, whether central-bank digital currency or just e-money guaranteed to be default-free (by being kept in a central bank) will enable contract-backed ('programmable') digital money (Henglein, 2022) with legally final delivery-versus-payment settlement within seconds of a trade and thus elimination of counterparty and settlement failure risk, but alignment of ESG taxonomies, securitization rulebook and settlement rules may also be required for the creation of a fully regulated, efficient secondary DLT-based capital market for VCM.

# 7    Conclusion

In this paper, we have investigated the general blockchain/DLT requirements given recent regulatory developments. We demonstrate how an artefact can be licensed as a fully compliant DLT-based trading and settlement system (DLT TSS) with positive network effects and the ability to deliver full market integrity, including motivations for exemptions to existing securities regulation according to the recent EU DLT Pilot Regime. DLT applications for carbon markets present significant benefit potential by providing transparency and traceability. However, the current markets lack integrity, and this lack of integrity is being exploited in permissionless blockchains to the extent that global verification bodies have suspended verification of the same. DLT with its associated technical and organizational innovations appear well-suited to deliver better solutions to capital markets by enabling a higher level of transparency, security, and legal certainty at substantially lower risk and cost. As the regulatory world cracks the door open to new technologies for improving security, transparency, investor protection, costs, and market efficiencies rather than instinctively associating them with anarcho-libertarian motives and as DLTR use cases present themselves as viable alternatives to legacy structures in the trading and settlement of securities, we believe the legacy capital market infrastructures will be challenged. The risks associated with this change are manageable, and the benefits appear attractive.